\documentclass[twocolumn,showpacs,preprintnumbers,amsmath,amssymb]{revtex4}
\usepackage{graphicx}
\usepackage{dcolumn}
\usepackage{color}
\usepackage{bm}
\begin{document}
\title {Dissipative Tunneling in 2 DEG: Effect of Magnetic Field, Impurity and Temperature.}
\vskip -0.5cm \author{ Malay Bandyopadhyay}
\affiliation{Nano Science Unit, S.N. Bose National Centre for Basic
Sciences,JD Block, Sector III, Salt Lake City, Kolkata 700098, India.}
\date{\today}
\begin{abstract}
We have studied the transport process in the two dimensional electron gas (2DEG) in presence of a magnetic field and a dissipative environment at temperature T. By means of imaginary time series functional integral method we calculate the decay rates at finite temperature and in the presence of dissipation. We have studied decay rates for wide range of temperatures ---- from the thermally activated region to very low temperature region where the system decays by quantum tunneling. We have shown that dissipation and impurity helps the tunneling. We have also shown that tunneling is strongly affected by the magnetic field. We have demonstrated analytical results for all the cases mentioned above.  
\end{abstract}
\maketitle
{\section {Introduction}}
The decay of a metastable state received an extensive study recently \cite{legg1,legg2} because of its important role played in physics, chemistry as well as transport in biomolecules \cite{vau,chan,marc}. The tunneling of two dimensional electron gas (2DEG) is a general phenomena and is widely encountered in mesoscopic systems \cite{sav,pao1,jain,pao2}, in semiconductor heterostructures \cite{butch,lee}, in quantum Hall systems \cite{zheng}. Here our focus is on tunneling of 2DEG in the presence of magnetic field, impurity potential and in a dissipative environment.\\
\noindent
In 2DEG the system can be thought of a collection of noninteracting charged particles moving in 2D. So one can visualize the system in question as if a charged particle of mass M moving in a metastable potential U(q) while coupled to a heat bath environment which is assumed here as a collection of harmonic oscillators. In macroscopic systems, the tunneling probability is strongly influenced by the interaction with the environment at finite temperature \cite{grab1,grab2}. The problem of tunneling in the presence of a coupling to many degrees of freedom such as phonons, magnons, gives rise to quantum friction which strongly affects the tunneling rate \cite{grab1,grab2,grab3,lang}. But all the above studies were in the absence of magnetic field. But here we shall study the tunneling process of 2DEG in presence of magnetic field. Besides dissipation we have special attention to the effects of impurity potential on the tunneling rate.\\
\noindent
Here we are actually discussing about the decay of a charged particle from a metastable potential well. Now at high temperatures the decay is thermally activated and the rate follows the famous Arrhenius law:
\begin{equation}
\Gamma_{cl} = f_{cl}exp(-\frac{v_b}{k_BT})
\end{equation}
where $V_b$ is the barrier height and $f_{cl}$ is the attempt frequency. As the temperature is lowered the classical escape rate decreases exponentially and it can decay only via quantum tunneling.\\
\noindent
\begin{figure}[h]
{\rotatebox{0}{\resizebox{5cm}{5cm}{\includegraphics{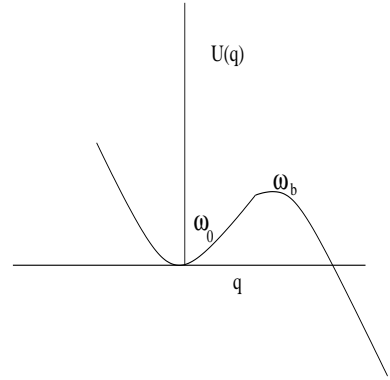}}}}
\caption{A metastable potential well.}
\end{figure}

Our method of analysis is based on pure thermodynamic equilibrium method which was pioneered by Langer \cite{lang}. In this approach the quantity of interest is the free energy of the metastable system. Because of states of lower energy on the other side of the barrier, the partition function can only b defined by means of an analytic continuation from a stable potential to the metastable situation depicted in Fig. 1. The procedure of analytic continuation leads to a unique Imaginary part of the free energy of the metastable state. This quantity is then related to the decay probability of the system. The thermal quantum rate of decay is related to $Im F$ by the following relation:
\begin{equation}
K = -\frac{2}{\hbar}Im F
\end{equation}
The method we employ to calculate the decay rate is a functional-integral approach which is convenient because it can easily take into account the dissipation as a nonlocal term. Beside this, we are dealing with a problem in a higher dimensional space: the particle moves in 2-D, and the effective dimension of the dissipative environment is infinite. The most convenient method for such problem is the path integral method which we use here. The Langer's method \cite{lang} for the calculation of the quantum decay rate was developed through the work of Miller, Stone, Coleman Callan \cite{miller}. They use the term ``bounce Euclidean action" $S_B$ and defined the quantum decay rate as
\begin{equation}
\Gamma = \omega_q exp(-\frac{S_B}{\hbar}),
\end{equation} 
where quantum mechanical preexponential factor $\omega_q$ is related with fluctuations about the bounce. Dissipation was first incorporated into the bounce technique by Caldeira and Leggett \cite{cal}. Also Grabert {\em et al} \cite{grab1,grab2,grab3} and Larkin {\em et al} \cite{larkin} used this method as an effective scheme to calculate decay rates in the entire range of temperature. In the present paper we extend these study by adding a couple of new features in addition to the usual dissipative effect:(a) the effect of the magnetic field, (b) the effect of the impurity potential. With the preceding background I organize the paper as follows. In $Sec. II$ I discuss about the free energy and Euclidean action of the system. We use the path integral method to reduce the infinite dimensional problem to an effective one dimensional problem. In $Sec. III$ I discuss about the thermodynamic method for the calculation of decay rates in the high temperature region. In $Sec. IV$ I talk about the quantum correction near the crossover region. Section V is our main focus. Below the crossover temperature the metastable state predominantly decay via quantum tunneling. Here we discuss about the effect of dissipation, impurity potential and magnetic field on the quantum tunneling rate of 2DEG in great details. Finally I summarize and conclude  the paper in $Sec. VI$.\\
{\section {Free Energy of the Damped 2DEG}}
I am specifically dealing with two dimensional electron gas (2 DEG) which confined in the x-y plane. The magnetic field is applied perpendicular to the x-y plane. Now following Caldeira and Leggett \cite{legg1,legg2} the system is assumed to be coupled linearly to its environment which can be represented as a collection of harmonic oscillators. Following Feynmann-Vernon \cite{feyn} one can integrates out all the environmental degrees of freedom and other irrelevant degrees of freedom. So the problem effectively comes out to be 1-D problem. Our starting point is the total Hamiltonian of the Global system i.e. the system and the reservoir which are coupled linearly:
\begin{equation}
{\cal H} = {\cal H_S} + {\cal H_B} + {\cal H_I},
\end{equation}
where subscripts $S$, $B$, and $I$ stand for `system',`bath' and `interaction' term respectively. 
\begin{eqnarray*}
{\cal H}& = & \sum_{l=x,y}\Big[\frac{1}{2M}\Big(p_{l} + \frac{e}{c}{A_{l}}({q_{l}})\Big)^{2}+U(q_{l})\Big]\\  & & +\sum_{j,l=x,y}\Big[\frac{1}{2m_j}p_{j,l}^{2} + \frac{1}{2}m_j\omega_{j}^{2}q_{j,l}^{2}\Big]\\ & & + \sum_{j,l=x,y}\frac{c_{j,l}^{2}}{2m_j\omega_{j}^{2}}q_{l}^{2} - \sum_{j,l=x,y}c_{j,l}q_{j,l}q_l.
\end{eqnarray*}
or,
\begin{eqnarray}
{\cal H} & = & \sum_{l=x,y}\Big[\frac{1}{2M}\Big(p_{l} + \frac{e}{c}{A_{l}}({q_{l}})\Big)^{2}+U(q_{l})\Big] \\ \nonumber & & +\sum_{j,l=x,y}\Big[\frac{1}{2m_j}p_{j,l}^{2} + \frac{1}{2}m_j\omega_{j}^{2}(q_{j,l}-\frac{c_{j,l}}{m_j\omega_{j}^{2}}q_{l})^{2}\Big],
\end{eqnarray}
where $q_l = (q_x,q_y)= \vec{q}$ are the co-ordinate of charged particle and $q_{j,l} = (q_{j,x},q_{j,y})$ are environmental degrees of freedom. Now all quantities characterizing the environment may be expressed in terms of the spectral density of bath oscillators
\begin{equation}
J_l(\omega) = \pi \sum_{j=1}{N} \frac{c_{j,l}^{2}}{2m_{j}\omega_{j}}\delta(\omega - \omega_{j}).
\end{equation}
Sometimes it is needed to introduce cut-off frequency $\omega_c$ \cite{pao1,pao2}. Thus
\begin{equation}
J_l(\omega) = \eta_{l}\omega^{s_l}exp(-\frac{\omega}{\omega_c}).
\end{equation}
In the present paper we follow the second definition of spectral density. The partition function of the charged Brownian particle can be written as a functional integral \cite{cal} over periodic paths where path probability is weighted according to the Euclidean action \cite{legg1,grab2,cal,hanggi}.
\begin{eqnarray}
S^{E}[\vec{q}(\tau),\dot{\vec{q}}(\tau)] & = & \int_{0}^{\hbar\beta}d\tau\Big[\frac{M}{2}\dot{\vec{q}}^{2}(\tau) + U(\vec{q}(\tau))\\ \nonumber & & - iM\omega_c(\vec{q}(\tau)\times\dot{\vec {q}}(\tau))_{z}\Big]\\ \nonumber & & + \sum_{j,l=x,y}\Big[\frac{1}{2m_j}p_{j,l}^{2} + \frac{1}{2}m_j\omega_{j}^{2}(q_{j,l}-\frac{c_{j,l}}{m_j\omega_{j}^{2}}q_{l})^{2}\Big]
\end{eqnarray}
The above action can be transformed into an effective action
\begin{eqnarray}
S^{E}[\vec{q}(\tau),\dot{\vec{q}}(\tau)] & = & \int_{0}^{\hbar\beta}d\tau\Big[\frac{M}{2}\dot{\vec{q}}^{2}(\tau) + U(\vec{q}(\tau))\\ \nonumber & & - iM\omega_c(\vec{q}(\tau)\times\dot{\vec {q}}(\tau))_{z}\Big]\\ \nonumber & & +\frac{1}{2}\int_{0}^{\hbar\beta}d\tau\int_{0}^{\hbar\beta}d\tau^{\prime}{\cal K}(\tau-\tau^{\prime})q(\tau)q(\tau^{\prime}),
\end{eqnarray}
where $\beta = \frac{1}{k_BT}$. The first term in Eq. (9) represents the conservative or reversible motion of the particle whereas the second term introduces the dissipation. The damping kernel ${\cal K}(\tau)$ may be expressed as a Fourier series
\begin{equation}
{\cal K}(\tau) = \frac{M}{\hbar\beta}\sum_{n=-\infty}^{+\infty}{\cal K}(\nu_n)exp(i\nu_n\tau),
\end{equation}
where Matsubara frequencies $\nu_n = \frac{2\pi nk_BT}{\hbar}$ and the Fourier coefficients are given by ${\cal K}(\nu_n) = \sum_{i}\frac{c_i^{2}}{m_i\omega_i^{2}}\frac{\nu_n^{2}}{\nu_n^{2}+\omega_i^{2}}$. The partition function is defined as
\begin{equation}
{\cal Z} = \int D[q]exp\Big[-\frac{1}{\hbar}S^{E}[q]\Big],
\end{equation}
where the functional integral is over all periodic paths with period $\hbar\beta$. The free energy is given by
\begin{equation}
F = -\frac{1}{\beta}ln{\cal Z}.
\end{equation}\\
{\section {Thermally Activated Decay}}
Temperature dependence of quantum decay rates in dissipative systems was first studied by Grabert {\em et al} \cite{grab1,grab2,grab3}. They define crossover temperature $T_0$ at which the transition between thermal hopping and quantum tunneling occurs. For temperature above $T_{0}$, the main contribution in the functional integral (11) arises from the vicinity of the time-independent trajectories $q(\tau) = 0$, where the particle sits on top of the potential barrier of the inverted potential, and $q(\tau) = q_b$, where it sits at the bottom of the well as shown in Fig. 2. a periodic path near $q(\tau) = 0$ can be expanded in terms of Matsubara frequencies $(\nu_n)$ as follows
\begin{equation}
x^{\prime}(\tau) = \sum_{n = -\infty}^{+\infty}X_n^{\prime}exp(i\nu_n\tau).
\end{equation}
\begin{figure}[h]
{\rotatebox{0}{\resizebox{5cm}{5cm}{\includegraphics{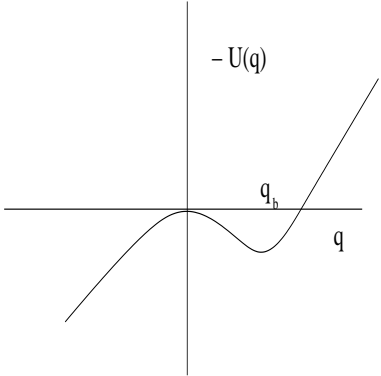}}}}
\caption{The inverted potential - U(q).}
\end{figure}

When we put Eq. (13) into Eq. (9) one obtains
\begin{equation}
S[X_n^{\prime}] = \frac{M\hbar\beta}{2}\sum_{n = -\infty}^{+\infty}\lambda_n^{0}X_n^{\prime}X_{-n}^{\prime},
\end{equation} 
where, $\lambda_n^{0} = \nu_{n}^{2}+\omega_0^{2}+2\omega_c\nu_n+|\nu_n|\hat{\gamma}(|\nu_n|)$, with well frequency $\omega_0 = [\frac{U^{\prime\prime}(0)}{M}]^{\frac{1}{2}}$ and cyclotron frequency $\omega_c = \frac{eB}{Mc}$. Now the partition function ${\cal Z}_0$ can be obtained by performing the Gaussian integrals over the amplitudes $X_n$. A periodic path around $q(\tau) = q_b$ can be written as
\begin{equation}
y^{\prime}(\tau) = q_b +\sum_{n = -\infty}^{+\infty}Y_n^{\prime}exp(i\nu_n\tau).
\end{equation}
Now the action becomes
\begin{equation}
S[Y_n^{\prime}] = \hbar\beta U_b + \frac{M\hbar\beta}{2}\sum_{n = -\infty}^{+\infty}\lambda_n^{b}Y_n^{\prime}Y_{-n}^{\prime},
\end{equation} 
where, $\lambda_n^{b} = \nu_{n}^{2}-\omega_b^{2}+2\omega_c\nu_n+|\nu_n|\hat{\gamma}(|\nu_n|)$, with frequency $\omega_b = [\frac{U^{\prime\prime}(b)}{M}]^{\frac{1}{2}}$. Since $\lambda_{0}^{b} = - \omega_{b}^{2}$; hence the integral over the amplitude $Y_{0}$ diverges. This leads to an imaginary part of free energy. The imaginary part of the free energy is given by
\begin{eqnarray}
Im F & = & -\frac{1}{\beta{\cal Z}_{0}}Im {\cal Z}_{b}\\ \nonumber
     & = & -\frac{1}{2\beta}\Big[\frac{D_0}{D_b}\Big]^{\frac{1}{2}}exp(-\beta U_b),
\end{eqnarray}
where,
\begin{equation}
D_0 = \prod_{n=-\infty}^{\infty}\lambda_n^{0};\ \ \ \ \    D_b = \prod_{n=-\infty}^{\infty}\lambda_n^{b}.
\end{equation}
Affleck \cite{aff} has shown that above $T_0$ the decay rate given by Eq. (2) is modified as follows
\begin{equation}
\Gamma = - \frac{2}{\hbar}\Big(\frac{T_0}{T}\Big)Im F.
\end{equation}
Now using the fact that $\lambda_0^{0}=\omega_0^{2}$ and $\lambda_0^{b}=-\omega_b^{2}$ and combining Eq. (17) with Eq. (19) we obtain
\begin{equation}
\Gamma = \frac{\omega_0}{2\pi}\frac{\omega_R}{\omega_b}\Big(\prod_{n=1}^{\infty}\frac{\nu_{n}^{2}+\omega_0^{2}+2\omega_c\nu_n+|\nu_n|\hat{\gamma}(|\nu_n|)}{\nu_{n}^{2}-\omega_b^{2}+2\omega_c\nu_n+|\nu_n|\hat{\gamma}(|\nu_n|)}\Big)exp(-\frac{U_b}{k_BT}),
\end{equation} 
where we made use the definition of crossover temperature $T_0 = \frac{\hbar\omega_R}{2\pi k_B}$ \cite{grab1,hanggi}. $\omega_R$ is the largest positive root of the equation $\omega_R^{2}+\omega_R(\hat{\gamma}(\omega_R)+2\omega_c) = \omega_b^{2}$. Now at high temperature $T>>T_0$ the product term in the bracket (...) becomes unity, thus we obtain
\begin {equation}
\Gamma_{cl} = \frac{\omega_0}{2\pi}\frac{\omega_R}{\omega_b}exp(-\frac{U_b}{k_BT}),
\end{equation}
which is the famous Arrhenius law. The influence of dissipation and magnetic field on the classical rate arises through memory renormalized barrier frequency $\omega_R$.\\
\noindent
As the temperature is lowered, the thermally activated decay rate is modified by the quantum correction factor. Quantum fluctuations enhances the decay probability because they increase the average energy of the Brownian charged particle. Now following Hanggi {\em et al} \cite{hanggi} we obtain the resulting leading quantum corrections are found to be given by the formula
\begin{equation}
f_q = exp\Big[\frac{\hbar^{2}}{24}\frac{(\omega_0^{2}+\omega_b^{2})}{(k_BT)^{2}}\Big]
\end{equation}
This $f_q$ is independent of dissipative and magnetic field mechanism. Let us now consider frequency independent damping case i.e. ohmic damping case. For the ohmic damping case the product in Eq. (20) can be evaluated in terms of $\Gamma$  functions as follows \cite{woly}
\begin{equation}
f_q = \frac{\Gamma(1-\frac{\lambda_b^{+}}{\nu})\Gamma(1-\frac{\lambda_b^{-}}{\nu})}{\Gamma(1-\frac{\lambda_0^{+}}{\nu})\Gamma(1-\frac{\lambda_0^{-}}{\nu})},
\end{equation}
where $\nu = \frac{2\pi k_BT}{\hbar}$ and
\begin{eqnarray}
\lambda_b^{\pm}& = &\frac{-(\gamma+2\omega_c)\pm\sqrt{(\gamma+2\omega_c)^{2}+4\omega_b^{2}}}{2};\\ \nonumber \lambda_0^{\pm}& = &\frac{-(\gamma+2\omega_c)\pm\sqrt{(\gamma+2\omega_c)^{2}-4\omega_0^{2}}}{2}.
\end{eqnarray}
For strongly damped system $\gamma>>\omega_0,\omega_b$ and $T>>T_0$ the ohmic quantum correction factor Eq. (23) becomes
\begin{equation}
f_q = exp\Big\lbrace\frac{T_0}{T}\Big[1+\frac{\omega_0^{2}}{\omega_b^{2}}\Big]\Big[\Psi\Big[1+4\alpha^{2}\frac{T_0}{T}\Big]-\Psi(1)\Big]\Big\rbrace,
\end{equation}
where $\Psi(x)$ is the digamma function and $\alpha = \frac{\gamma+2\omega_c}{2\omega_b}$. On the other hand, for intermediate temperatures $T_0<<T<<4\alpha^{2}T_0$ we obtain the quantum correction factor \cite{grab1}
\begin{equation}
f_q = \Big(\frac{4\alpha^{2}T_0}{T}\Big)^{\frac{(\omega_b^{2}+\omega_0^{2})T_0}{\omega_b^{2}T}}.
\end{equation}
Now in the third case where $\gamma<<\omega_0,\omega_b$ we have \cite{grab1}
\begin{equation}
f_q = \frac{\omega_b}{\omega_0}\frac{sinh\Big(\frac{\hbar\omega_0}{2k_BT}\Big)}{sin\Big(\frac{\hbar\omega_b}{2k_BT}\Big)}exp[D\alpha+O(\alpha^{2})],
\end{equation}
where $\alpha = \frac{\gamma+2\omega_c}{2\omega_b}$ and
\begin{equation}
D = \frac{\omega_b}{\nu}\Big[\Psi(1+\frac{\omega_b}{\nu})+\Psi(1-\frac{\omega_b}{\nu})-\Psi(1+i\frac{\omega_0}{\nu})-\Psi(1-i\frac{\omega_0}{\nu})\Big].
\end{equation}\\
{\section {NEAR CROSSOVER REGION}}
In the above section we have seen that the quantum corrections become very much important as we approach the crossover temperature $T_0$. In the vicinity of $T_0$ the above mentioned semiclassical approximation fails because the eigenvalue $\lambda_1^{b}=\lambda_{-1}^{b}=\nu^{2}-\omega_b^{2}+(2\omega_c+\hat{\gamma}(\nu))\nu$ vanishes for $T=T_0$. Hence the Gaussian integral over the amplitude $Y_1,Y_{-1}$ diverges. Now to regularize this divergent integral I add the higher order terms in amplitudes $Y_1,Y_{-1}$ \cite{grab1}. I expand the potential U(q) about the barrier top $U(q) = U_b - \frac{1}{2}M\omega_b^{2}(q-q_b)^2 + \sum_{k=3}^{\infty}\frac{1}{k}Mc_k(q-q_b)^k$. Now following Grabert {\em et al} I obtain the quantum correction factor in the crossover region as follows:
\begin{equation}
f_q = \Big(\frac{\lambda_1^{0}}{\Lambda_1}\Big)f_R,
\end{equation}
where 
\begin{equation}
f_R = \prod_{n=2}^{\infty}\Big(\frac{\lambda_n^{0}}{\lambda_n^{b}}\Big)\nonumber,
\end{equation}
$\lambda_n^{0} = \nu_n^{2}+\omega_0^{2}+(2\omega_c+\hat{\gamma}(\nu_n))\nu_n$,\\$\lambda_n^{b} = \nu_n^{2}-\omega_b^{2}+(2\omega_c+\hat{\gamma}(\nu_n))\nu_n$,\\$\Lambda_1 = \Big[\Big(\frac{\pi M\beta}{2B}\Big)^{\frac{1}{2}}erfc\Big[\lambda_1^{b}\Big(\frac{M\beta}{2B}\Big)^{\frac{1}{2}}\Big]exp\Big[(\lambda_1^{b})^{2}(\frac{M\beta}{2B})\Big]\Big]^{-1}$ and $B = \frac{4c_3^{2}}{\omega_b^{2}+\omega_c^{2}}-\frac{2c_2^{2}}{\lambda_2^{b}} + 3c_4$.\\
A more detailed analysis on the crossover region has been discussed by many others \cite{grab1,weiss,grab2,grab3}, although they did not consider the magnetic field effect which I have considered here. Now I switch over to our main focus i.e. the quantum tunneling region.\\
{\section {QUANTUM TUNNELING REGION}}
Below the crossover temperature $T_0$, the tunneling through the barrier becomes more probable and the decay rate is mainly given by the tunneling rate. In this region the most probable escape path is the bounce trajectory $q_B(\tau)$. For temperature below $T_0$ the imaginary part of the free energy comes from the so-called bounce trajectory \cite{woly}. The decay rate is given by \cite{grab1,grab2,grab3}:
\begin{equation}
\Gamma = \Big(\frac{S_0}{2\pi\hbar}\Big)\Big(\frac{D_0}{D_B^{\prime}}\Big)^{\frac{1}{2}}exp\Big(-\frac{S_B}{\hbar}\Big),
\end{equation}
where $S_B$ is the action (9) evaluated along the bounce trajectory $q_B(\tau)$, $S_0$ is the zero-mode normalization factor
\begin{equation}
S_0 = M\int_{0}^{\hbar\beta}d\tau \dot{q_B}^{2}(\tau),
\end{equation}
$D_0$ is the product of eigenvalues of fluctuation modes about the metastable minimum, and $D_B^{\prime}$ is the product of eigenvalues of fluctuation modes about the bounce trajectory with the zero eigenvalue omitted.\\
\noindent
In my case I am interested in the 2D electron gas tunneling out of the metastable state $q_x = 0$ in the x direction. The electron gas is moving in the x-y plane and magnetic field is applied perpendicular to the x-y plane. The 2DEG is coupled with a dissipative environment. so far, we have considered arbitrary forms of the metastable potential. Now I consider the practically important case of a cubic potential, $U_1(q_x) = \frac{1}{2}M\omega_x^{2}\Big(q_x^{2}-\frac{q_x^{3}}{q_0}\Big)$,which is relevant for SQUID or a current-biased Josephson junction. I also approximate the impurity potential in the y direction by the harmonic potential. Thus $U(q_x,q_y) = \frac{1}{2}M\omega_x^{2}\Big(q_x^{2}-\frac{q_x^{3}}{q_0}\Big) + \frac{1}{2}M\omega_y^{2}q_y^{2}$. For my convenience I use the notation $\vec{x} = (x,y)$ instead of $\vec{q}_l = (q_x,q_y)$ . The tunneling is described by the Euclidean action 
\begin{eqnarray}
S^{E}& = & \int_{0}^{\hbar\beta}d\tau\Big[\frac{1}{2}M(\dot{x}^{2}+\dot{y}^{2})+iM\omega_c\dot{x}y+U_1(x)\Big]\\ \nonumber & & +\frac{1}{2}M\omega_y^{2}y^{2} +\sum_{j}\Big[\frac{1}{2m_j}\vec{p}_{j}^{2} + \frac{1}{2}m_j\omega_{j}^{2}(\vec{q}_{j}-\frac{c_{j}}{m_j\omega_{j}^{2}}\vec{x})^{2}\Big],
\end{eqnarray}
where I have used Landau gauge. After the tunneling the other degrees of freedoms, $y$ and ${q_j}$ can take any allowed values. Hence one can easily integrate out the environmental degrees of freedom ${q_j}$ and $y$ coordinates. Before doing all these integration over ${q_j}$ and $y$ coordinates, we perform a Fourier transformation in the time interval $[0,\hbar\beta]$:
\begin{eqnarray}
\Big(x(\tau),y(\tau)\Big)& = & \frac{1}{\hbar\beta}\sum_{n=-\infty}^{+\infty}(x_n,y_n)e^{i\nu_n\tau} \\ 
q_j &=& \frac{1}{\hbar\beta}\sum_{n=-\infty}^{+\infty} q_{j,n}e^{i\nu_n\tau}
\end{eqnarray}
Now the action can be rewritten in terms of Fourier components as follows:
\begin{eqnarray*}
S^{E}& = & \frac{1}{\hbar\beta}\sum_{n=-\infty}^{+\infty}\Big[\frac{1}{2}M\nu_n^{2}x_nx_{- n}+\frac{1}{2}(M\nu_n^{2} + M\omega_y^{2})y_ny_{- n}\Big]\\
\nonumber 
& & +U_1(x_n) + \frac{1}{\hbar\beta}\sum_{n=-\infty}^{+\infty}M\omega_cx_ny_{- n}\nu_{- n}\\
\nonumber 
& & +\frac{1}{\hbar\beta}\sum_{n=-\infty}^{+\infty}\sum_{j}\Big[\frac{1}{2}m_j\nu_n^{2}\vec{q}_{j,n}\vec{q}_{j,- n}\\ 
\nonumber 
& & + \frac{1}{2}m_j\omega_j^{2}\Big(\vec{q}_{j,n}-\frac{c_j}{m_j\omega_j^{2}}\vec{x}_{n}\Big)\Big(\vec{q}_{j,- n}-\frac{c_j}{m_j\omega_j^{2}}\vec{x}_{- n}\Big)\Big],
\end{eqnarray*}
with $U_1(x_n) = \hbar\beta\int_{0}^{\hbar\beta}d\tau U(x(\tau).$ After doing integration over the environmental degrees of freedom $q_{j,n}$, I obtain
\begin{eqnarray*}
S^{E}& = & \frac{1}{\hbar\beta}\sum_{n=-\infty}^{+\infty}\frac{1}{2}(M\nu_n^{2}+\xi_{x,n})x_nx_{- n}\\ & & +\frac{1}{\hbar\beta}\sum_{n=-\infty}^{+\infty}\frac{1}{2}(M\nu_n^{2} + M\omega_y^{2}+\xi_{y,n})y_ny_{- n}\\
& & +U_1(x_n) + \frac{1}{\hbar\beta}\sum_{n=-\infty}^{+\infty}M\omega_cx_ny_{- n}\nu_{- n},
\end{eqnarray*}
with $ \xi_{l,n} = \frac{1}{\pi}\int_{0}^{\infty}d\omega \frac{J_{l}(\omega)}{\omega}\frac{2\nu_n^{2}}{\nu_n^{2}+\omega^{2}};\ \ \ \ \ l=x,y.$ Here we are considering tunneling along x-direction, hence I can integrate out the y variables. After doing the integration over y variables I obtain the effective one dimensional action as follows: 
\begin{eqnarray}
S^{E}& = & \frac{1}{\hbar\beta}\sum_{n=-\infty}^{+\infty}\frac{1}{2}(M\nu_n^{2}+\xi_{x,n})x_nx_{- n} + U_1(x_n)\\ \nonumber & & -\frac{1}{\hbar\beta}\sum_{n=-\infty}^{+\infty}\frac{(M\omega_c)^{2}\nu_n\nu_{- n}}{M\nu_n^2+M\omega_y^{2}+\xi_{y,n}}x_nx_{- n}
\end{eqnarray}
The above effective 1D action can be rewritten as shown below:
\begin{eqnarray}
S_{eff}^{E}[x(\tau)] & = & \int_{0}^{\hbar\beta}d\tau \Big[\frac{1}{2}M\dot{x}^{2}(\tau)+U_1(x(\tau))\Big]\\ \nonumber & &
+\frac{1}{2}\int_{0}^{\hbar\beta}d\tau\int_{0}^{\hbar\beta}d\tau^{\prime}\Big[{\cal K}(\tau-\tau^{\prime})+g(\tau-\tau^{\prime})\Big]\\ \nonumber & &\ \ \ \ \
\Big(x(\tau)-x(\tau^{\prime})\Big)^{2},
\end{eqnarray}
where the normal damping kernel is given by
\begin{equation}
{\cal K}(\tau) = \frac{1}{\pi}\int_{0}d\omega J(\omega) \frac{cosh[\omega(\frac{\hbar\beta}{2}-\tau)]}{sinh[\frac{\omega\hbar\beta}{2}]},
\end{equation}
and the anomalous damping kernel is given by
\begin{eqnarray}
g(\tau) & = & -\frac{1}{\hbar\beta}\sum_{n=-\infty}^{+\infty}\frac{(M\omega_c)^{2}\nu_n^{2}}{M\nu_n^2+M\omega_y^{2}+\xi_{y,n}}e^{i\nu_n\tau}\\ \nonumber & = &
-\frac{1}{\hbar\beta}(M\omega_c)^{2}\Big[\sum_{n=-\infty}^{+\infty}e^{i\nu_n\tau}\\ \nonumber & & -\sum_{n=-\infty}^{+\infty}\frac{M\omega_y^{2}+\xi_{y,n}}{M\nu_n^2+M\omega_y^{2}+\xi_{y,n}}e^{i\nu_n\tau}\Big].
\end{eqnarray}
The first term is the delta function and has no contribution to the semiclassical action, thus finally I obtain
\begin{equation}
g(\tau)  =  \frac{1}{\hbar\beta}(M\omega_c)^{2}\sum_{n=-\infty}^{+\infty}\frac{M\omega_y^{2}+\xi_{y,n}}{M\nu_n^2+M\omega_y^{2}+\xi_{y,n}}e^{i\nu_n\tau}.
\end{equation}
To derive the normal damping kernel I use the identity \cite{weiss} $\sum_n\frac{\nu_n^2}{\nu_n^2+\omega^2}e^{i\nu_n\tau}=\frac{\hbar\beta}{2}\frac{cosh[\omega(\frac{\hbar\beta}{2}-\tau)]}{sinh[\frac{\omega\hbar\beta}{2}]}.$
Thus we have obtained an effective one dimensional problem. The physics due to the influence of the dissipative environment and magnetic field is underlined in the normal and anomalous damping kernel respectively. Now I shall discuss different limiting cases.

{(i)\bf \underline {Impurity zero and no dissipation}}

In this limiting condition the normal damping kernel ${\cal K}(\tau) = 0$ and the anomalous damping kernel $g(\tau) = \frac{(M\omega_c)^2}{\hbar\beta}$. Thus the effective Euclidean action becomes
\begin{eqnarray}
S_{eff}^{E}& = &\int_{0}^{\hbar\beta}d\tau\Big[\frac{1}{2}M\dot{x}^2(\tau)+U_1(x(\tau))\Big]\\ \nonumber & & +\frac{(M\omega_c)^2}{\hbar\beta}\int_{0}^{\hbar\beta}d\tau\int_{0}^{\hbar\beta}d\tau^{\prime}[x(\tau)-x(\tau^{\prime})]^2.
\end{eqnarray}
$S_{eff}^{E}$ depends on magnetic field through $\omega_c$ and this dependence is proportional to $B^2$. Thus for a strong magnetic field the tunneling rate [Eq. (30)] vanishes at very low temperature. At zero temperature the effective action Eq. (40) can be written as 
\begin{equation}
S_{eff}^{E} = \int_{0}^{\hbar\beta}d\tau\Big[\frac{1}{2}M\dot{x}^2(\tau)+U_1(x(\tau))+\frac{1}{2}M\omega_c^{2}x^{2}(\tau)\Big].
\end{equation}
Equation (41) shows that the magnetic field renormalizes the potential $U_1(x(\tau)$ such that the local minima at $ x=0$ is now more stable. Because the renormalized potential is now a double well potential and thus the second local minima may be lower than at $ x = 0$. 

{(ii)\bf \underline {Finite impurity and no dissipation}}

Since there is no dissipation, hence the normal damping kernel ${\cal K}(\tau)=0$ and the anomalous damping kernel is given by 
\begin{eqnarray*}
g(\tau) & = & -\frac{1}{\hbar\beta}(M\omega_c)^{2}\sum_{n=-\infty}^{+\infty}\frac{\nu_n^{2}}{M\nu_n^2+M\omega_y^{2}+\xi_{y,n}}e^{i\nu_n\tau}\\ & = &-\frac{1}{\hbar\beta}(M\omega_c)^{2}\frac{1}{M}\sum_{n=-\infty}^{+\infty}\frac{\nu_n^{2}}{\nu_n^2+\omega_y^{2}}\\ & = & -\frac{1}{\hbar\beta}(M\omega_c)^{2}\frac{1}{M}\frac{\hbar\beta}{2}\frac{cosh[\omega_y(\frac{\hbar\beta}{2}-\tau)]}{sinh[\frac{\omega_y\hbar\beta}{2}]}   
\end{eqnarray*}
Thus the effective euclidean action becomes
\begin{eqnarray}
S_{eff}^{E}& = &\int_{0}^{\hbar\beta}d\tau \Big[\frac{1}{2}M\dot{x}^2(\tau)+U_1(x(\tau))\Big]\\ \nonumber & & -\frac{1}{4M}(M\omega_c)^{2}\int_{0}^{\hbar\beta}d\tau\int_{0}^{\hbar\beta}d\tau^{\prime}\frac{cosh[\omega_y(\frac{\hbar\beta}{2}-|\tau-\tau^{\prime}|)]}{sinh[\frac{\omega_y\hbar\beta}{2}]}\\ \nonumber & & \ \ \ \ \ [x(\tau)-x(\tau^{\prime})]^2
\end{eqnarray}
Now one can easily be able to understand that the tunneling rate is finite for any value of magnetic field. Thus the impurity acts in opposite way as that of magnetic field. So impurity suppresses the effect of magnetic field and it actually enhances the tunneling rate.

{(i)\bf \underline {Impurity zero and finite dissipation}}

I set the impurity potential i.e. $\omega_y^{2} = 0$. We know that the quantum tunneling predominates at very low temperature, i.e. at very large imaginary time limit $(\hbar\beta\rightarrow\infty)$. So we have to take large time limit values of the normal and anomalous damping kernel \cite{pao1,pao2}. In the large $\tau$ limit the normal damping kernel is given by \cite{pao1,pao2} ${\cal K}(\tau) = \frac{1}{\pi}\eta \frac{1}{\tau^{s+1}}$ and the anomalous damping kernel becomes $g(\tau) = \frac{(M\omega_c)^2}{2\pi M}\frac{1}{\eta^{\prime}}\frac{1}{\tau^{2-s+1}}$, where $\eta^{\prime} = \frac{\eta}{2M}\frac{2}{\pi}\int_0^{\infty}dz\frac{z^{s-1}}{z^2+1}$. Thus the effective action for the ohmic damping (s=1) case becomes
\begin{eqnarray*}
S_{eff}^{E}& = &\int_{0}^{\infty}d\tau\Big[\frac{1}{2}M\dot{x}^2(\tau)+U_1(x(\tau))\Big]\\ \nonumber & & 
+\frac{1}{2}\int_{0}^{\infty}d\tau\int_{0}^{\infty}d\tau^{\prime}\Big[\frac{1}{\pi}\eta\frac{1}{|(\tau-\tau^{\prime})|^{1+1}}\\ \nonumber & & +\frac{(M\omega_c)^2}{2\pi M}\frac{1}{|(\tau-\tau^{\prime})|^{2}}\frac{1}{\frac{\eta}{2M}\frac{2}{\pi}\int_0^{\infty}\frac{dz}{z^2+1}}\Big][x(\tau)-x(\tau^{\prime})]^2
, 
\end{eqnarray*}
or,
\begin{eqnarray*}
S_{eff}^{E} & = & \int_{0}^{\infty}d\tau\Big[\frac{1}{2}M\dot{x}^2(\tau)+U_1(x(\tau))\Big]\\ \nonumber & & +\frac{\Big(\eta+\frac{(M\omega_c)^{2}}{\eta}\Big)}{2\pi}\int_{0}^{\infty}d\tau\int_{0}^{\infty}d\tau^{\prime}\frac{1}{|(\tau-\tau^{\prime})|^{2}}\\& &[x(\tau)-x(\tau^{\prime})]^2.
\end{eqnarray*}
Finally I obtain
\begin{eqnarray}
S_{eff}^{E} & = & \int_{0}^{\infty}d\tau\Big[\frac{1}{2}M\dot{x}^2(\tau)+U_1(x(\tau))\Big]\\ \nonumber & &+ \frac{\eta_{eff}}{2\pi}\int_{0}^{\infty}d\tau\int_{0}^{\infty}d\tau^{\prime}\frac{1}{|\tau-\tau^{\prime}|^2}[x(\tau)-x(\tau^{\prime})]^2,
\end{eqnarray}
where $\eta_{eff} = \eta + \frac{(M\omega_c)^2}{\eta}$. From the Eq. (43) it is well understood that the effect of dissipation is to suppress the effect of magnetic field. Because of the factor $\frac{1}{|\tau-\tau^{\prime}|^2}$ we should get finite tunneling at any amount of magnetic field strength. So dissipative environment helps in tunneling of 2DEG in presence of magnetic field.

{(i)\bf \underline {Finite impurity and finite dissipation}}

Since both impurity and dissipation helps in tunneling of 2DEG in presence of magnetic field, so the total effect of them will be the same. The effective action is
\begin{eqnarray*}
S_{eff}^{E} & = & \int_{0}^{\hbar\beta}d\tau\Big[\frac{1}{2}M\dot{x}^2(\tau)+U_1(x(\tau))\Big]\\ \nonumber & &+\frac{1}{2}\int_{0}^{\hbar\beta}d\tau\int_{0}^{\hbar\beta}d\tau^{\prime}[{\cal K}(|\tau-\tau^{\prime}|)+g(\tau-\tau^{\prime})]\\ & & [x(\tau)-x(\tau^{\prime})]^2.
\end{eqnarray*}
Now in case of strong impurity the anomalous damping kernel is negligible compared to the normal damping kernel, because in this case the anomalous superohmic kernel decays rapidly compare to ohmic normal damping kernel. Hence the effective action becomes:
\begin{eqnarray}
S_{eff}^{E} & = & \int_{0}^{\hbar\beta}d\tau\Big[\frac{1}{2}M\dot{x}^2(\tau)+U_1(x(\tau))\Big]\\ \nonumber & & + \frac{\eta}{2\pi}\int_{0}^{\infty}d\tau\int_{0}^{\infty}d\tau^{\prime}\frac{1}{|\tau-\tau^{\prime}|^2}[x(\tau)-x(\tau^{\prime})]^2.
\end{eqnarray}
Equation (44) suggests that dirty sample enhances the tunneling rate of 2DEG in presence of magnetic field. On the other hand for strong damping case one can neglect the superohmic normal damping kernel compared to ohmic anomalous damping kernel. Thus the effective action now reads as:
\begin{eqnarray}
S_{eff}^{E} & = & \int_{0}^{\hbar\beta}d\tau\Big[\frac{1}{2}M\dot{x}^2(\tau)+U_1(x(\tau))\Big]\\ \nonumber & & + \frac{(M\omega_c)^2}{2\pi\eta}\int_{0}^{\infty}d\tau\int_{0}^{\infty}d\tau^{\prime}\frac{1}{|\tau-\tau^{\prime}|^2}[x(\tau)-x(\tau^{\prime})]^2.
\end{eqnarray}
From the above expression it is well understood that dissipation opposes the magnetic field effect.

{(i)\bf \underline {Very strong magnetic field}}

Under very strong magnetic field the kinetic energy of the electron gas freezes. Thus one obtain the effective action as follows
\begin{eqnarray*}
S_{eff}^{E} & = & \int_{0}^{\hbar\beta}d\tau\Big[iM\omega_c(\dot{x}(\tau)y(\tau)) + U_1(x(\tau))+\frac{1}{2}M\omega_y^2 y^2]\\ \nonumber & & +\frac{1}{2}\int_{0}^{\infty}d\tau\int_{0}^{\infty}d\tau^{\prime}{\cal K}(\tau -\tau^{\prime})[x(\tau)-x(\tau^{\prime})]^2.
\end{eqnarray*}
Following Jain {\em et al} \cite{jain} I obtain 
\begin{eqnarray*}
S_{cl}^{E} & = & -i\int_{0}^{x_t}dx iM\omega_c(\dot{x}(\tau)y(\tau))\\ & & + \frac{1}{2}\int_{0}^{\infty}d\tau\int_{0}^{\infty}d\tau^{\prime}\frac{1}{|\tau-\tau^{\prime}|^2}[x_{cl}(\tau)-x_{cl}(\tau^{\prime})]^2.
\end{eqnarray*} 
Finally I obtain
\begin{eqnarray}
S_{cl} & = & M\omega_c\int_{0}^{x_t}dx \Big[\frac{2U_1(x)}{M\omega_y^2}\Big]^{\frac{1}{2}} \\ \nonumber & & + \frac{1}{2}\int_{0}^{\infty}d\tau\int_{0}^{\infty}d\tau^{\prime}\frac{1}{|\tau-\tau^{\prime}|^2}[x_{cl}(\tau)-x_{cl}(\tau^{\prime})]^2.
\end{eqnarray}
Thus as we increase the impurity potential along y-direction $S_{cl}$ decreases and the tunneling rate of the 2DEG along x-direction increases.\\

{\section {CONCLUSION AND SUMMARY}}

In this paper I have presented the imaginary time functional integral approach to reaction rates calculation. I have provided a complete description of the decay of a metastable state extending from the classical regime to quantum region. I have covered a wide range of temperature region --- extending from the high temperature thermally activated region to a very low temperature quantum tunneling  regime. It has been seen that well above the crossover temperature $T_0$, the decay rate follows the famous Arrhenius law. Here the preexponential factor is affected by the damping and magnetic field through the renormalized frequency $\omega_R$. Now as we approach to the crossover temperature the quantum correction factor to the decay rate becomes very much important. Now well below the crossover temperature the system decays predominantly by quantum tunneling process. I have shown that at zero temperature the tunneling rate of 2DEG from the metastable state vanishes in a strong magnetic field and in absence of both dissipation and impurity. The effect of impurity potential and the dissipation have been discussed in great details in Sec. IV. Both the impurity and dissipation subdue the effect of magnetic field and enhances the tunneling rate. I have discussed this tunneling rate in context of a very relevant cubic potential. So my results are also applicable to the physically interesting problem of quantum tunneling in SQUID or current-biased Josephson junctions.
\section*{Acknowledgments}   
I wish to express my gratitude to Prof. Sushanta Dattagupta for his constant encouragement. Financial support from CSIR is gratefully acknowledged. \\

\end{document}